\documentclass[a4paper]{article}

\usepackage{INTERSPEECH2021}

\usepackage[utf8]{inputenc}
\usepackage{bbm}
\usepackage{multirow}
\usepackage{makecell}
\usepackage{cite}
\usepackage{amssymb,subfigure}
\usepackage{placeins}
\usepackage{babel,blindtext}
\usepackage{caption}
\usepackage{tikz}
\usepackage{adjustbox}
\usepackage{verbatim}
\usepackage{hyperref}
\usepackage{xurl}
\usepackage{cleveref}
 
\usepackage[T1]{fontenc}
\usepackage{amsmath,graphicx}
\title{Fre-GAN: Adversarial Frequency-consistent Audio Synthesis}
\name{Ji-Hoon Kim$^1$, Sang-Hoon Lee$^2$, Ji-Hyun Lee$^1$, Seong-Whan Lee$^{1,2}$ \thanks{This work was supported by Institute of Information \& communications Technology Planning \& Evaluation (IITP) grant funded by the Korea government (MSIT) (No. 2019-0-00079, Department of Artificial Intelligence, Korea University), and the Magellan Division of Netmarble Corporation.}}
\address{
  $^1$Department of Artificial Intelligence, Korea University, Seoul, Korea\\
  $^2$Department of Brain and Cognitive Engineering, Korea University, Seoul, Korea}
\email{\{jihoon\_kim, sh\_lee, jihyun-lee, sw.lee\}@korea.ac.kr}

\begin{document}

\maketitle
\begin{abstract}
Although recent works on neural vocoder have improved the quality of synthesized audio, there still exists a gap between generated and ground-truth audio in frequency space. This difference leads to spectral artifacts such as hissing noise or reverberation, and thus degrades the sample quality. In this paper, we propose Fre-GAN which achieves frequency-consistent audio synthesis with highly improved generation quality. Specifically, we first present resolution-connected generator and resolution-wise discriminators, which help learn various scales of spectral distributions over multiple frequency bands. Additionally, to reproduce high-frequency components accurately, we leverage discrete wavelet transform in the discriminators. From our experiments, Fre-GAN achieves high-fidelity waveform generation with a gap of only $0.03$ MOS compared to ground-truth audio while outperforming standard models in quality.
\end{abstract}

\section{Introduction}
Deep generative models have revolutionized neural vocoder that aims to transform intermediate acoustic features such as mel-spectrogram into intelligible human speech. Especially, autoregressive models\cite{oord2016wavenet,kalchbrenner2018efficient} have shown exceptional perfor-mance in terms of quality and replaced the role of conventional approaches\cite{griffin1984signal,morise2016world}. Nevertheless, they suffer from slow inference speed owing to their autoregressive nature. To address the structural limitation of them, flow-based vocoders are proposed\cite{kim2018flowavenet,prenger2019waveglow,yoon2020audio}. While they can produce natural waveform in real-time because of their ability to convert noise sequence into raw waveform in parallel, they require a heavy computation due to the complex architecture\cite{lee2020multi}.

The other approach is based on Generative Adversarial  Networks (GANs)\cite{kumar2019melgan,yamamoto2020parallel,yang2020vocgan,kong2020hifi,jang2020universal}. MelGAN\cite{kumar2019melgan} adopts Multi-Scale Discriminator (MSD)\cite{wang2018high} operating on multiple 
scales of waveforms modulated by Average Pooling (AP). The MSD has been proven to be advantageous for capturing consecutive patterns and long-term dependencies of audio \cite{yang2020vocgan,kong2020hifi,jang2020universal}. Parallel WaveGAN\cite{yamamoto2020parallel} proposes multi-resolution spectrogram loss which helps to stabilize adversarial training and improve the generation quality. Recently, HiFi-GAN\cite{kong2020hifi} identifies the periodic patterns of audio through Multi-Period Discriminator (MPD) and synthesizes high-fidelity audio. It further improves the audio quality by applying a Multi-Receptive field Fusion (MRF) module in the generator which observes patterns of diverse lengths in parallel. The model outperforms the autoregressive\cite{oord2016wavenet} and flow-based vocoder\cite{prenger2019waveglow} in terms of both quality and inference speed. Despite the recent advances, there still exists a gap between synthesized and ground-truth audio in frequency space. This gap leads to spectral artifacts such as hissing noise or robotic sound, because audio is made of a complicated mixture of various frequencies.  
\begin{figure}[!t]
    \centering
    \hspace{-0.04\columnwidth} 
    \subfigure[Average Pooling]{
    \centering
    \includegraphics[width=.48\columnwidth]{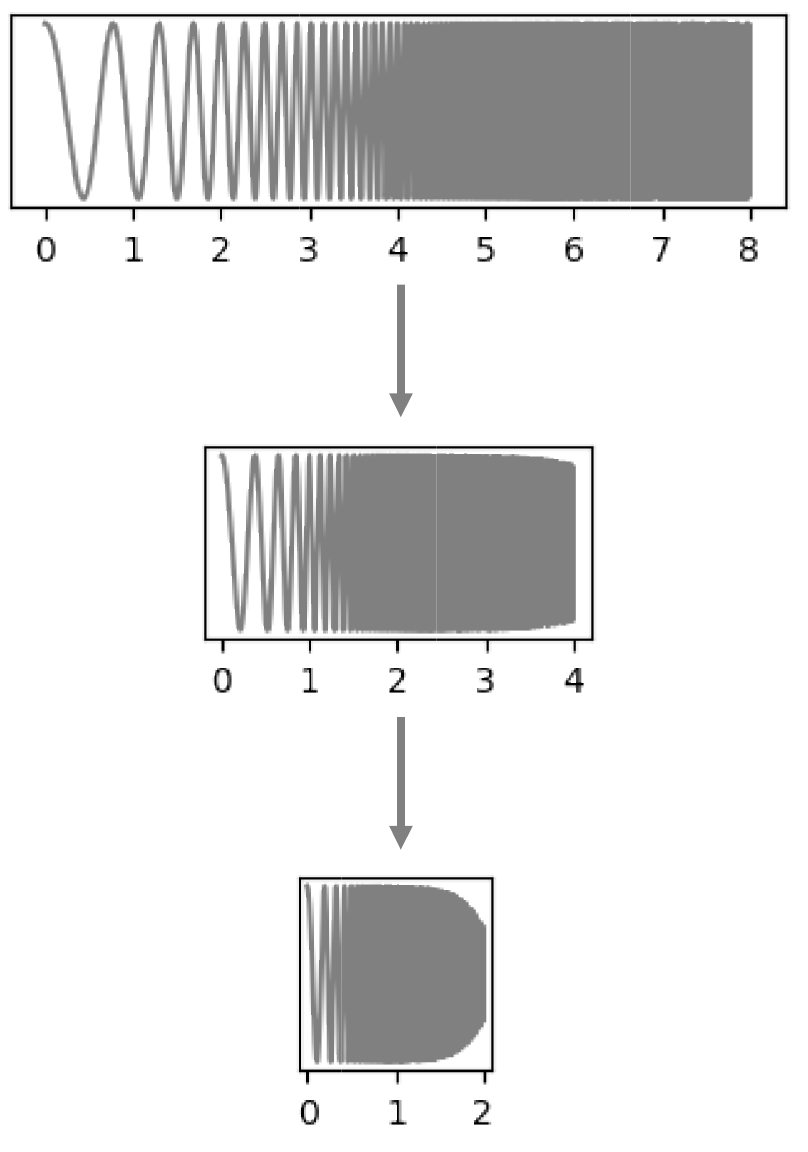}
    \label{fig:down-avg}}
    \hspace{-0.02\columnwidth} 
    \subfigure[Discrete Wavelet Transform]{
    \includegraphics[width=.48\columnwidth]{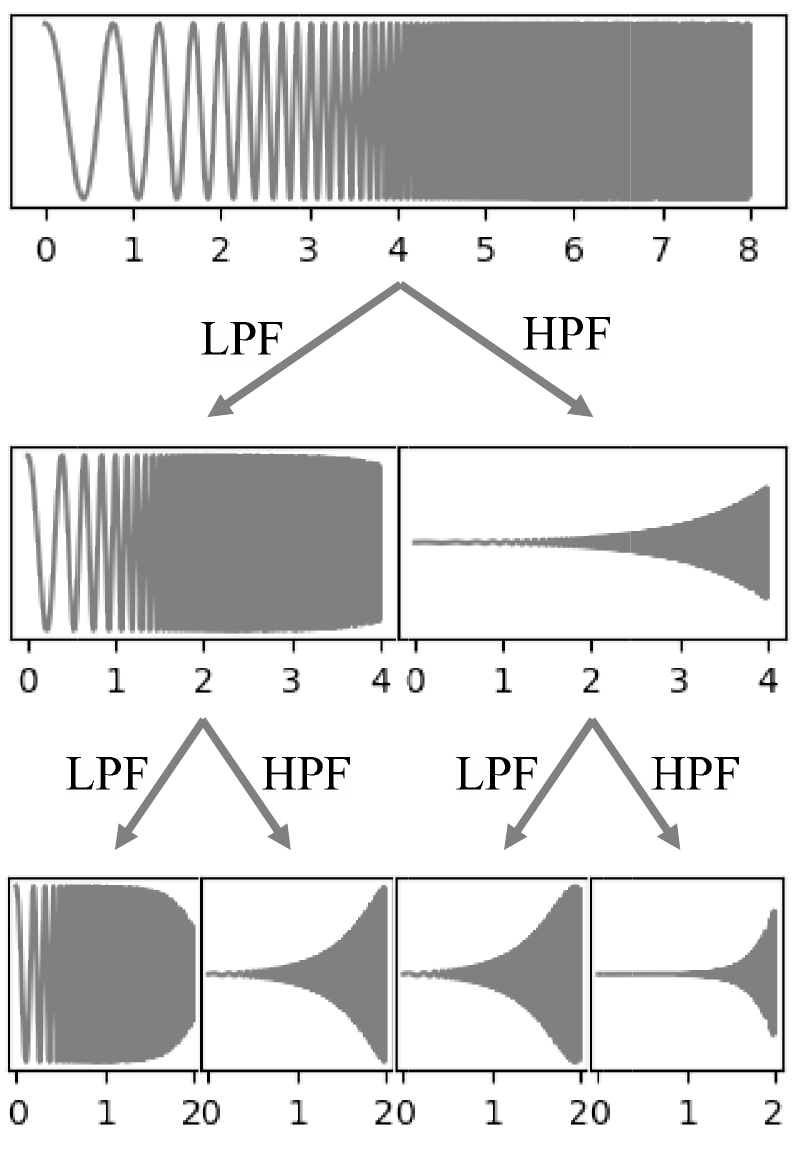}
    \label{fig:down-dwt}}  
    \hspace{-0.04\columnwidth} 
    \caption{Comparison of Average Pooling (AP) and Discrete Wavelet Transform (DWT). Here, LPF and HPF refer to Low-Pass and High-Pass Filter, respectively. In this example, an up-chirp signal whose frequency increases from 0 Hz at time $t=0$ to 150 Hz at time $t=8$ is downsampled by AP and DWT.}
    \label{fig:downs}
\end{figure}

In this paper, we propose Fre-GAN which synthesizes frequency-consistent audio on par with ground-truth audio. Fre-GAN employs resolution-connected generator and resolution-wise discriminators to learn various levels of spectral distributions over multiple frequency bands. The generator upsamples and sums multiple waveforms at different resolutions. Each waveform is adversarially evaluated in the corresponding resolution layers of the discriminators. To further facilitate the training of discriminators, we provide the downsampled audio to each resolution layer of the discriminators\cite{gulrajani2017improved,miyato2018spectral,karras2020analyzing}. Based on this architecture, we use Discrete Wavelet Transform (DWT) as the downsampling method. While the conventional downsampling method, i.e. AP, washes the high-frequency components away, DWT guarantees that all the information can be kept due to its biorthogonal property. In Fig.~\ref{fig:downs}, we provide evidence for the above statement. Unlike AP, DWT can safely deconstruct the signal into low-frequency and high-frequency sub-bands without losing high-frequency contents. In the experiments, the effectiveness of Fre-GAN is demonstrated on various metrics.

\begin{figure*}[!t]
    \centering
    \subfigure[RCG]{
    \includegraphics[width=.6\columnwidth]{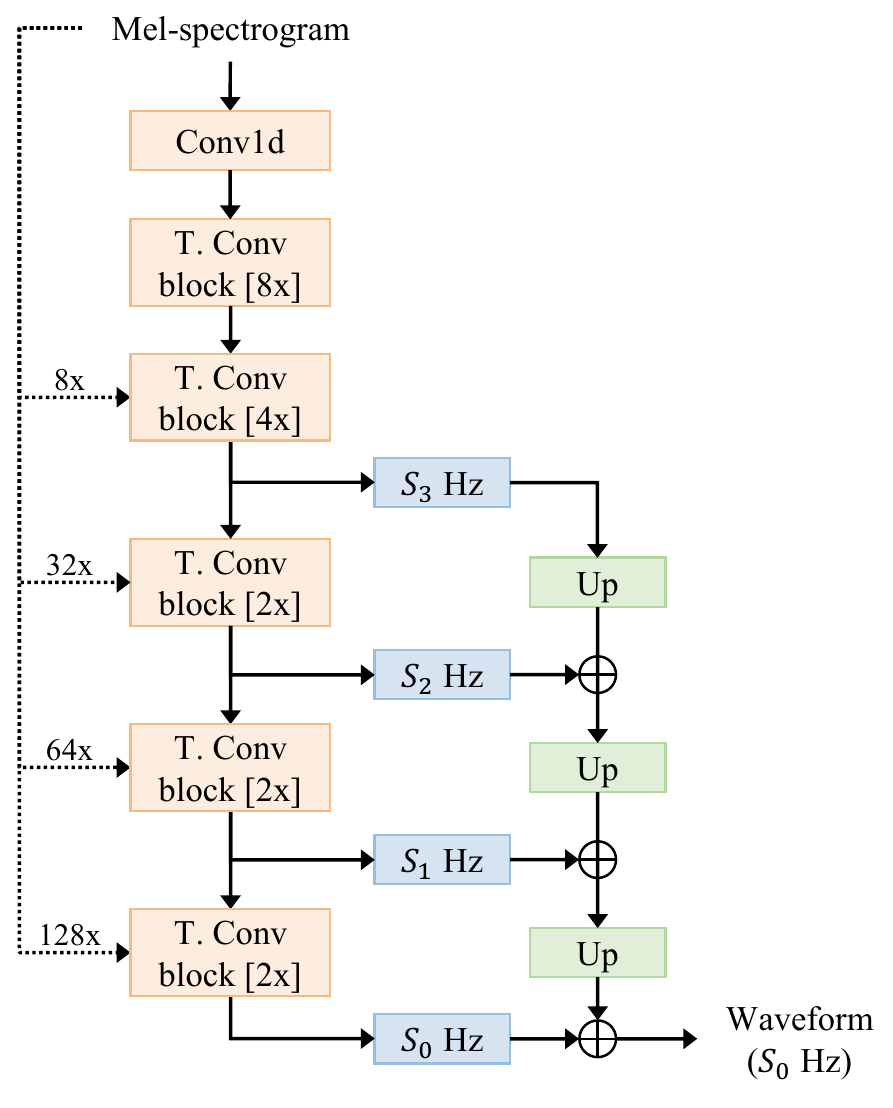}
    \label{fig:gen}
    }
    \subfigure[RPD]{
    \includegraphics[width=.6\columnwidth]{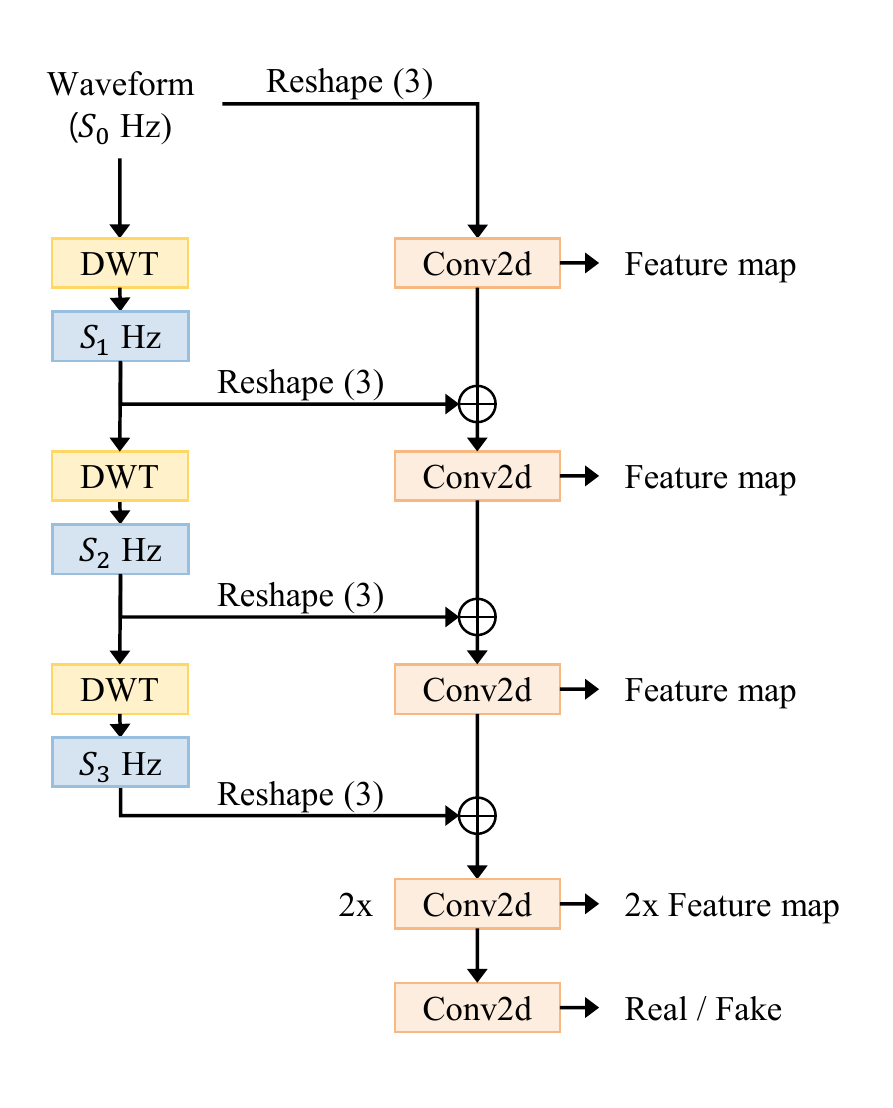}
    \label{fig:MPD}
    } 
    \subfigure[RSD]{
    \includegraphics[width=.6\columnwidth]{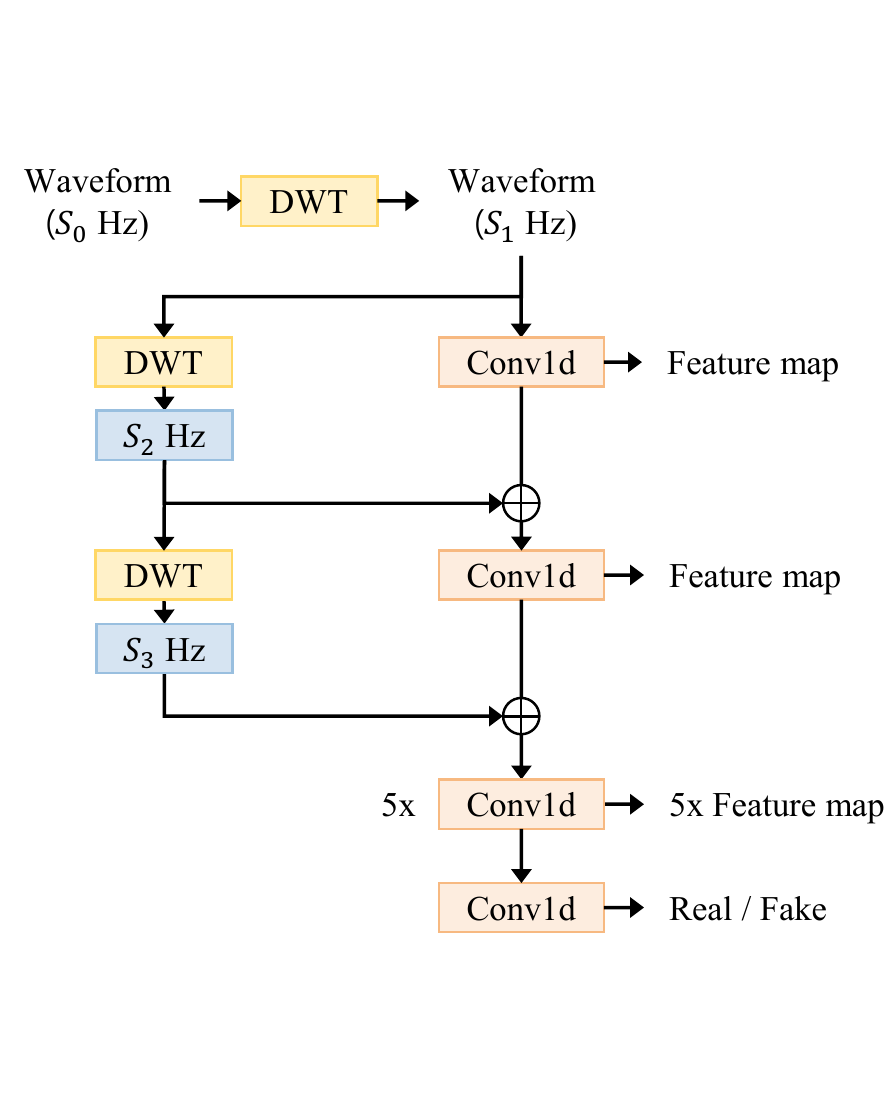}
    \label{fig:MSD}
    }
    \caption{Fre-GAN network architecture. (a) The RCG. (b) The second sub-discriminator of RPD. (c) The second sub-discriminator of RSD. T. Conv block denotes transposed convolution block. In this research, $256\times$ upsampling is conducted in $5$ stages of $8\times$, $4\times$, $2\times$, $2\times$, and $2\times$. UP denotes nearest neighbor upsampler which consists of nearest neighbor interpolation and $1\times1$ convolution. Reshape $(p)$ refers to reshaping the raw audio from 1d into 2d data with period $p$. }
\label{model}
\end{figure*}

\section{Fre-GAN}
\subsection{Resolution-connected Generator}
\label{RCG}
Fre-GAN generator takes a mel-spectrogram as input and upsamples it through transposed convolution blocks until the temporal resolution of the output sequence matches that of the raw waveform. The transposed convolution blocks consist of transposed convolutional layer followed by MRF module proposed in HiFi-GAN\cite{kong2020hifi} and leaky-relu activation.

Inspired by StyleGAN2\cite{karras2020analyzing}, we adopt skip-connections to the generator and call it Resolution-Connected Generator (RCG). The RCG upsamples and sums top-$K$ waveform outputs corresponding to different resolutions, as illustrated in Fig.~\ref{fig:gen}. To upsample lower scale waveforms, we use Nearest Neighbor (NN) upsampler\cite{pons2020upsampling} which has been proven to alleviate tonal artifacts caused by transposed convolutions\cite{odena2016deconvolution}. In addition, we directly condition input mel-spectrogram to each top-$K$ transposed convolution block. This allows the multiple waveforms to be consistent with the input mel-spectrogram. In this research, we set $K$ to four. 

We investigate that the RCG structure has several benefits. Firstly, it captures various levels of spectral distributions by explicitly summing multiple waveforms at different resolutions. This instigates the model to effectively learn acoustic properties across multiple frequency bands. Secondly, the RCG is trained through progressive learning\cite{karras2017progressive}. The training of RCG starts by focusing on low resolution and then progressively shifts attention to higher resolutions, as we will validate it in Sec.~\ref{sec:ablation}. This allows the model to first discover the easier coarse structure and then shift its focus to learn increasingly finer details, instead of learning all scales at once. By gradually increasing the resolution, we can speed up and greatly stabilize adversarial training.

\begin{table*} [!ht]
    \caption{Evaluation results. The MOS are presented with $95\%$ confidence intervals. Higher is better for MOS and speed, and lower is better for the other metrics. Speed of $n$ kHz means that the model can synthesize $n\times1000$ audio samples per second. The numbers in () denote real-time factor. Values in bold represent the best results for each metric.}
    \centering
    \begin{tabular}{l|c|c|c|c|r r|r r} \Xhline{3\arrayrulewidth}
        \textbf{Model} &\textbf{MOS $(\uparrow)$} &\textbf{MCD$_{13}$ $(\downarrow)$} &\textbf{RMSE$_{f0}$ $(\downarrow)$} &\textbf{FDSD $(\downarrow)$} &\multicolumn{2}{c|}{\textbf{Speed on CPU} $(\uparrow)$} &\multicolumn{2}{c}{\textbf{Speed on GPU $(\uparrow)$}} \\ \hline
            Ground Truth &$4.40\pm{0.04}$ &\textendash &\textendash &\textendash &\multicolumn{2}{c|}{\textendash} &\multicolumn{2}{c}{\textendash}\\ \hline 
            WaveNet &$4.20\pm{0.06}$ &$2.293$ &$43.347$ &$0.725$  &\multicolumn{2}{c|}{\textendash}  &$0.12$\hspace{-0.03\columnwidth} &$(\times 0.005)$\hspace{0.03\columnwidth}   \\
            WaveGlow &$3.94\pm{0.07}$ &$2.048$ &$40.463$ &$0.542$ &$15.83$\hspace{-0.03\columnwidth} &$(\times 0.72)$\hspace{0.03\columnwidth} &$319$\hspace{-0.03\columnwidth} &$(\times 14.47)$\hspace{0.03\columnwidth}     \\ 
            HiFi-GAN V1 &$4.30\pm{0.05}$ &$1.231$ &$39.947$ &$0.190$ &$53.20$\hspace{-0.03\columnwidth} &$(\times 2.41)$\hspace{0.03\columnwidth} &$2,351$\hspace{-0.03\columnwidth} &$(\times 106.62)$\hspace{0.03\columnwidth}    \\ 
            HiFi-GAN V2 &$4.19\pm{0.05}$ &$1.606$ &$40.258$ &$0.282$ &$\textbf{215.43}$\hspace{-0.03\columnwidth} &$(\times \textbf{9.77})$\hspace{0.03\columnwidth} & $\textbf{10,730}$\hspace{-0.03\columnwidth} &$(\times \textbf{486.62})$\hspace{0.03\columnwidth}  \\   \hline 
            Fre-GAN V1 &$\textbf{4.37}\pm\textbf{0.04}$ &$\textbf{1.060}$ &$\textbf{38.339}$ &$\textbf{0.150}$ &$60.72$\hspace{-0.03\columnwidth} &$(\times 2.75)$\hspace{0.03\columnwidth} &$2,284$\hspace{-0.03\columnwidth} &$(\times 103.58)$\hspace{0.03\columnwidth} \\ 
            Fre-GAN V2 &$4.28\pm{0.05}$ &${1.308}$ &${38.843}$ &${0.205}$ &$192.07$\hspace{-0.03\columnwidth} &$(\times 8.71)$\hspace{0.03\columnwidth} &$10,458$\hspace{-0.03\columnwidth} &$(\times 474.29)$\hspace{0.03\columnwidth} \\ 
       \Xhline{3\arrayrulewidth}
    \end{tabular}
    \label{MOS}
\end{table*}

\subsection{Resolution-wise Discriminators}
Fre-GAN employs two discriminators: Resolution-wise MPD (RPD) and Resolution-wise MSD (RSD), whose architectures are drawn from HiFi-GAN\cite{kong2020hifi}. RPD comprises five sub-discriminators each of which accepts specific periodic parts of input audio; the period is given by $p\in\{2,3,5,7,11\}$ to avoid overlaps\cite{kong2020hifi}. To be specific, the input audio of length $T$ is first reshaped to 2d data of height $T/p$ and width $p$, and then applied to 2d convolutions. Whereas RSD consists of three sub-discriminators operating on different input scales: raw audio, $2\times$ downsampled audio, and $4\times$ downsampled audio. Note that RPD captures periodic patterns of audio and RSD observes consecutive patterns of audio. 

By setting the resolutions of convolution layers in each sub-discriminator to match that of top-$K$ waveforms in the RCG, we encourage a specific layer in each sub-discriminator to evaluate the waveform of the corresponding resolution. This resolution-wise adversarial evaluation provokes the RCG to learn the mapping from input mel-spectrogram to audio at various scales. Moreover, we provide downsampled audio to the corresponding resolution layer of each sub-discriminator, as shown in Fig.~\ref{fig:MPD} and~\ref{fig:MSD}. As proven in recent works\cite{gulrajani2017improved,miyato2018spectral,karras2020analyzing}, this residual connection facilitates the discriminators training and improves the sample quality. Based on this structure, we use Discrete Wavelet Transform (DWT) to downsample audio without losing any information.

\subsection{Discrete Wavelet Transform}
\label{DWT}
In previous works using MSD, they have used AP to downsample raw audio\cite{kumar2019melgan, yang2020vocgan,kong2020hifi,jang2020universal}. However, the AP ignores the sampling theorem\cite{zhang2019making}, and high-frequency contents are aliased and become invalid\cite{chen2020ssd}. This makes the generator lack the incentives from the MSD to learn high-frequency components, resulting in a spectral distortion in high-frequency bands. 

To alleviate high-frequency loss, we replace problematic AP with Discrete Wavelet Transform (DWT)\cite{daubechies1988orthonormal} as our down-sampling method. The DWT is an efficient but effective way of downsampling non-stationary signals into several frequency sub-bands. In 1d DWT, the signal is convolved by two filters: low-pass filter $(g)$ and high-pass filter $(h)$. According to Nyquist's rule\cite{nyquist1928certain}, half the samples of the convolution results can be discarded since half the frequencies of the signal are removed. This gives two $2\times$ downsampled signals representing low-frequency and high-frequency components, respectively. They can be further decomposed by DWT as follows:

\begin{equation}
    y_{low}[n]=\sum_{k}x[k]g[2n-k]
\end{equation}
\begin{equation}
    y_{high}[n]=\sum_{k}x[k]h[2n-k]
\end{equation}
where $y_{low}[n]$ and $y_{high}[n]$ are subsequent outputs of $g$ and $h
$, respectively. $n$ and $k$ denote levels of DWT and index of signal $x$, respectively. Due to the biorthogonal property of DWT, the signal can be deconstructed safely without information loss. 

After each level of DWT, all the frequency sub-bands are channel-wise concatenated and passed to convolutional layers\cite{suk2011subject,lee2018high}. When all the sub-bands are taken into account, Fre-GAN can avoid information loss, especially in high frequencies. In our implementation, Daubechies1 wavelet\cite{daubechies1988orthonormal} is adopted.

\subsection{Training Objectives}
To train Fre-GAN, we use the least-squares GAN objective because of its training stability\cite{mao2017least}. The training objectives for the discriminators and generator are defined as:
\begin{equation}
    \begin{split}
    \hspace{-0.1cm}
    \mathcal{L}_D=
         &\sum_{n=0}^{4}\mathbb{E}[\lVert D_{n}^{P}(x)-1\rVert_2+\lVert D_{n}^{P}(\hat{x})\rVert_2]\\
         &+\sum_{m=0}^{2}\mathbb{E}[\lVert D_{m}^{S}(\phi^{m}(x)-1)\rVert_2+\lVert D_{m}^{S}(\phi^{m}(\hat{x}))\rVert_2]
    \end{split}
\end{equation}
\begin{equation}
    \begin{split}
    \mathcal{L}_G=&\sum_{n=0}^{4}\mathbb{E}[\lVert D_{n}^{P}(\hat{x})-1\rVert_2+\lambda_{fm}\mathcal{L}_{fm}(G;D_{n}^{P})] \\
        &+\sum_{m=0}^{2}\mathbb{E}[\lVert D_{m}^{S}(\hat{x})-1\rVert_2+\lambda_{fm}\mathcal{L}_{fm}(G;D_{m}^{S})] \\
        &+\lambda_{mel}\mathcal{L}_{mel}(G)
    \end{split}
\end{equation}
where $x$ and $\hat{x}$ denote ground-truth and generated audio, respectively. $D^{P}$ and $D^{S}$ are RPD and RSD, respectively. $\phi^{m}$ represents $m$-level DWT. In the experiments, we set $\lambda_{fm}=2$ and $\lambda_{mel}=45$ which balance the adversarial losses, the feature matching loss ($\mathcal{L}_{fm}$), and the mel-spectrogram loss ($\mathcal{L}_{mel}$) defined as follows: 
\begin{equation}
\begin{split}
    \mathcal{L}_{fm}(G;D_k)= &\mathbb{E}\bigg[\sum_{i=0}^{T-1}\frac{1}{N_i}\lVert
    D_{k}^{(i)}(x)-D_{k}^{(i)}(\hat{x})\rVert_1\bigg]
\end{split}
\end{equation}
\begin{equation}
    \mathcal{L}_{mel}(G)=\mathbb{E}\bigg[\lVert\psi(x)-\psi(\hat{x})\rVert_1\bigg]
\end{equation}
Here, $T$ denotes the number of layers in the discriminator. $D_{k}^{(i)}$ is the $i^{th}$ layer feature map of the $k^{th}$ sub-discriminator, $N_i$ is the number of units in each layer, and $\psi$ is the STFT function to convert raw audio into the corresponding mel-spectrogram. 

The feature matching loss minimizes $L1$ distance between the discriminator feature maps of real and generated audio\cite{larsen2016autoencoding}. As it was successfully adopted to neural vocoder\cite{kumar2019melgan,kong2020hifi}, we use it as an auxiliary loss to improve the training efficiency. In addition, we add mel-spectrogram loss to further improve the sample quality and training stability. It is the reconstruction loss that minimizes $L1$ distance between the mel-spectrogram of synthesized audio and that of ground-truth audio. Referring to previous works\cite{isola2017image,yamamoto2020parallel}, applying reconstruction loss to GANs training helps to generate realistic results and stabilize the adversarial training process from the early stages.

\section{Experimental Results}
We conducted experiments on LJSpeech dataset at a sampling rate of $22,050$ Hz. The dataset contains $13,100$ audio samples of a single English speaker, and we randomly split the dataset into train $(80\%)$, validation $(10\%)$, and test $(10\%)$ sets. Fre-GAN was compared against several neural vocoders trained on the same dataset: the popular open-source implementation of mixture of logistics WaveNet\cite{wavenet:impl}, and the official implementation of WaveGlow\cite{waveglow:impl} and HiFi-GAN\cite{HiFi:impl}. All the models were trained for $3100$ epochs.

Similar to HiFi-GAN\cite{kong2020hifi}, we conducted experiments based on two variations of the generator: V1, V2 with the same discriminator configuration. Each variation resembles that of HiFi-GAN but we set the kernel sizes of transposed convolutions to $[16,8,4,4,4]$ and dilation rates of MRF to $[[1,1],[3,1],[5,1],[7,1]\times3]$. $80$ bands mel-spectrogram was transformed with $1024$ of window size, $256$ of hop size, and $1024$ points of Fourier transform. We used AdamW optimizer \cite{loshchilov2017decoupled} with $\beta_{1}=0.8$, $\beta_{2}=0.999$, $16$ of batch size, and followed the same learning rate schedule in \cite{kong2020hifi}. The synthesized audio samples are available at \url{https://prml-lab-speech-team.github.io/demo/FreGAN} 
\begin{figure*}[!t]
    \centering
    \subfigure[Fre-GAN V1]{
    \includegraphics[height=2.25cm]{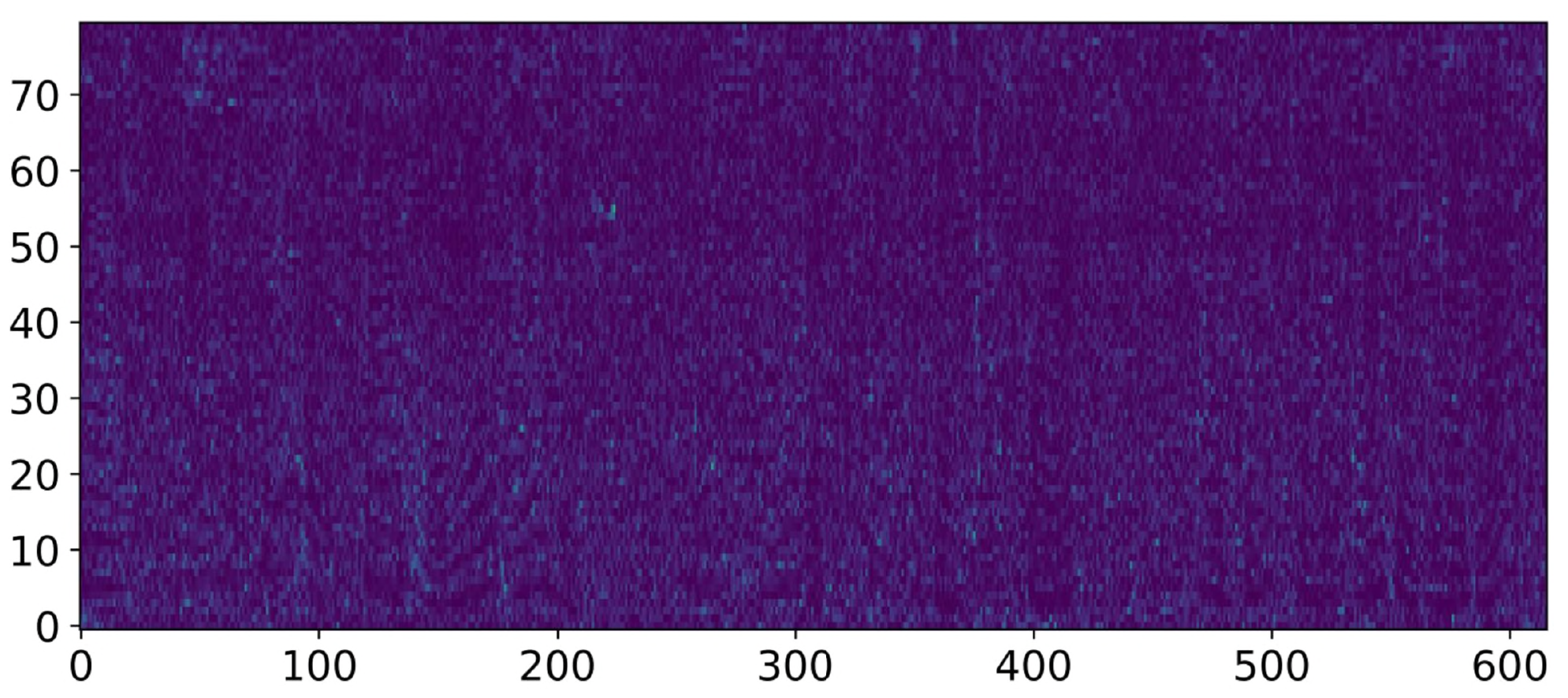}
    \label{fig:dif-frev1}}
    \subfigure[HiFi-GAN V1]{
    \includegraphics[height=2.25cm]{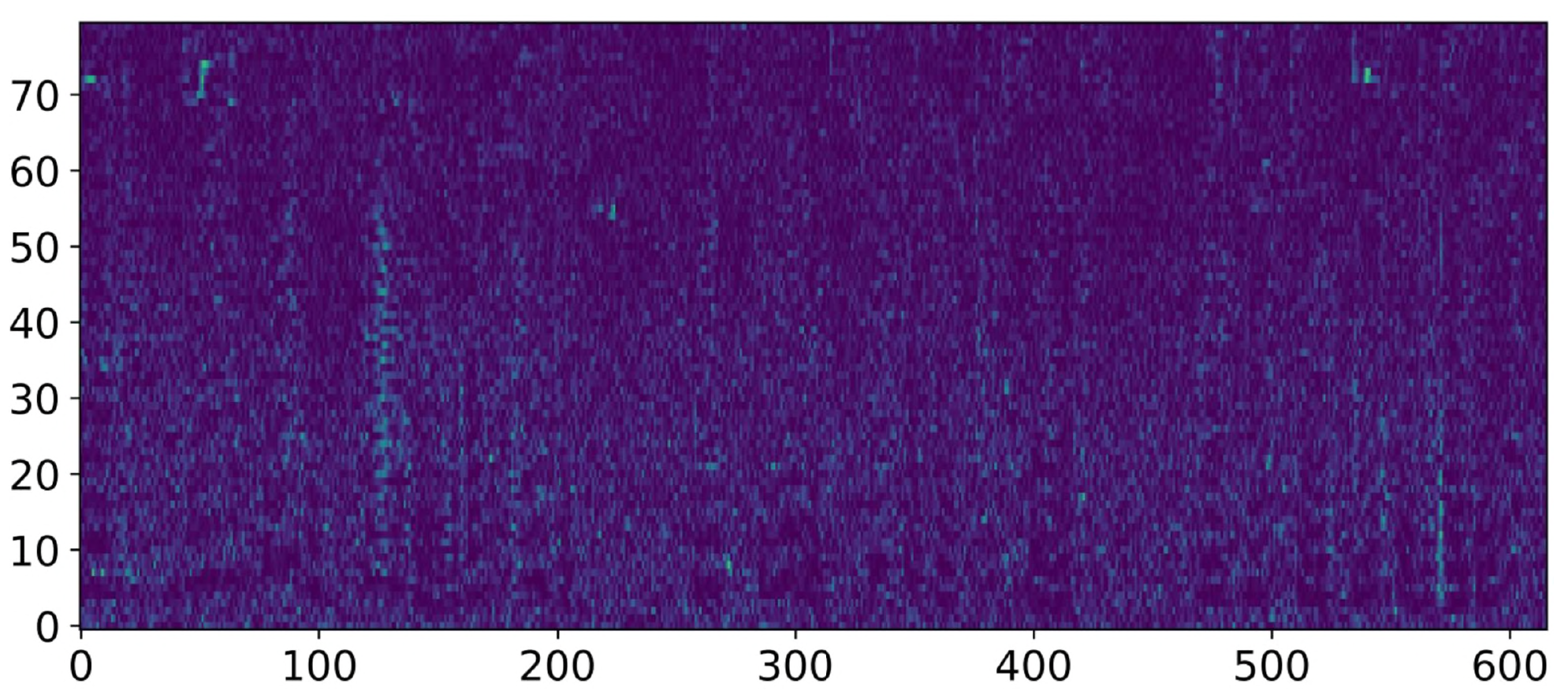}
    \label{fig:dif-hiv1}}
    \subfigure[WaveNet]{
    \includegraphics[height=2.25cm]{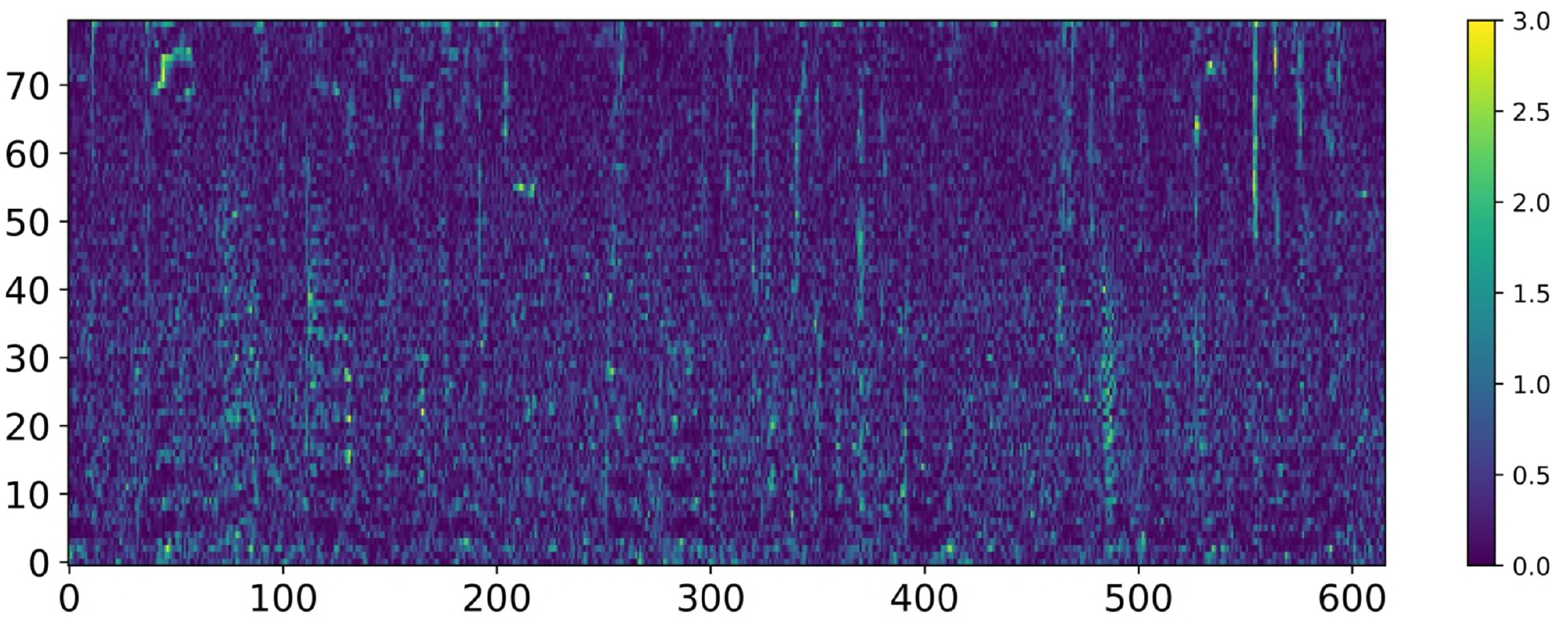}
    \label{fig:dif-wavenet}}
    
    \subfigure[Fre-GAN V2]{
    \includegraphics[height=2.25cm]{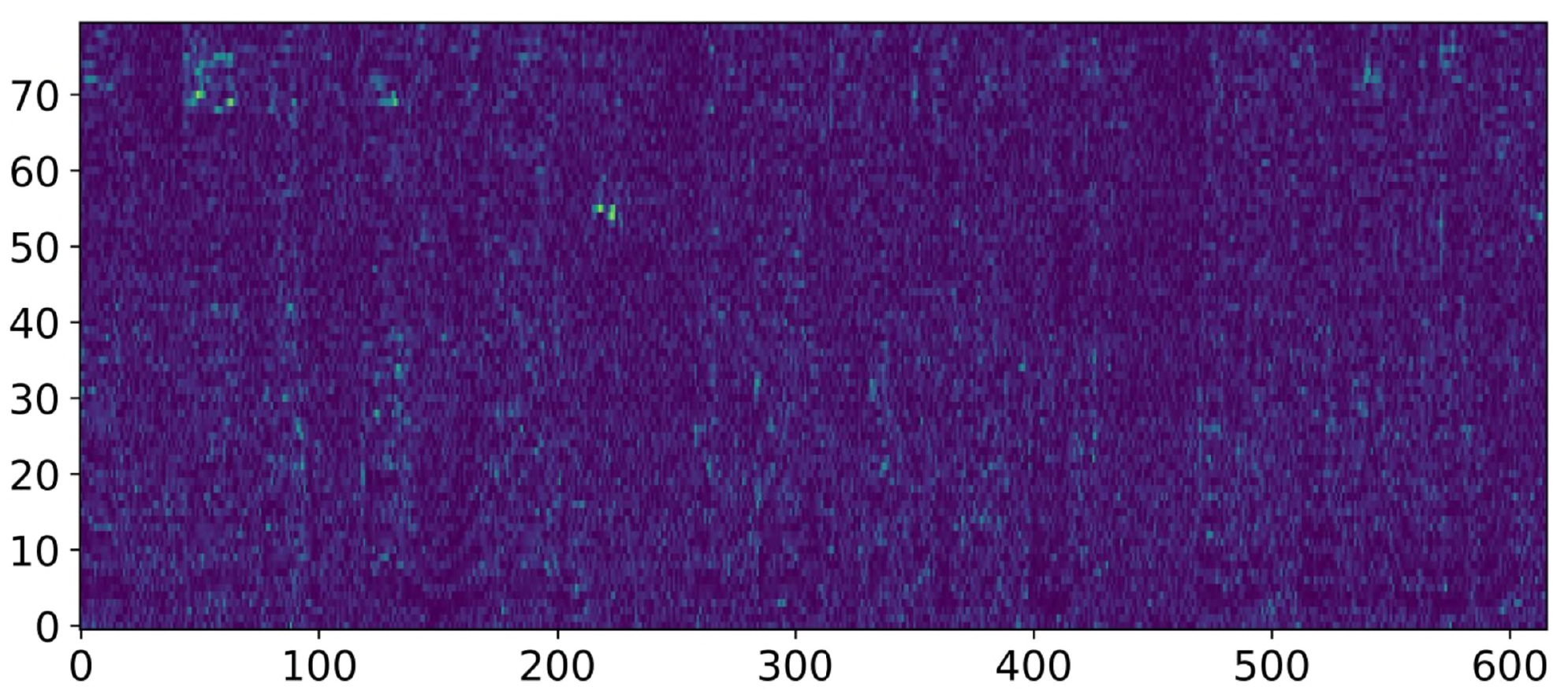}
    \label{fig:dif-frev2}}
    \subfigure[HiFi-GAN V2]{
    \includegraphics[height=2.25cm]{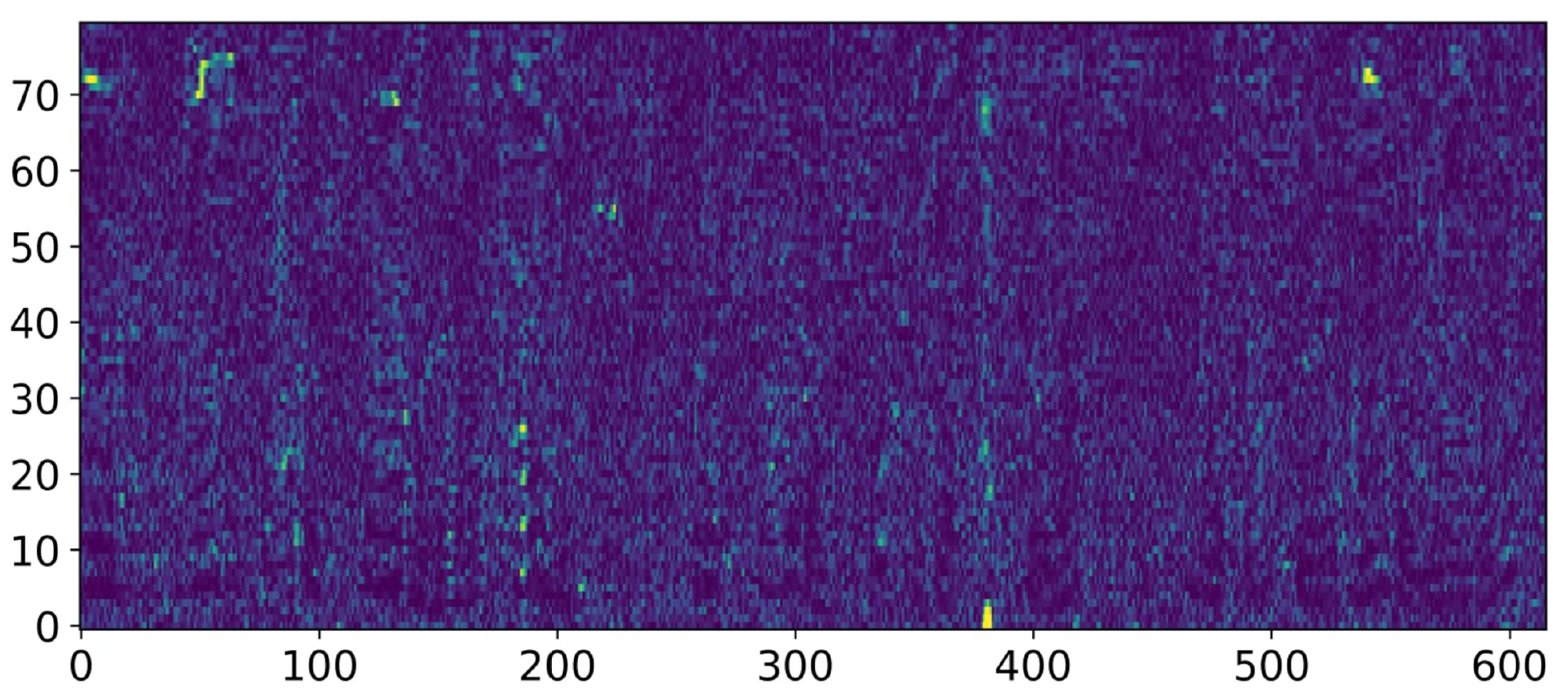}
    \label{fig:dif-hiv2}}
    \subfigure[WaveGlow]{
    \includegraphics[height=2.25cm]{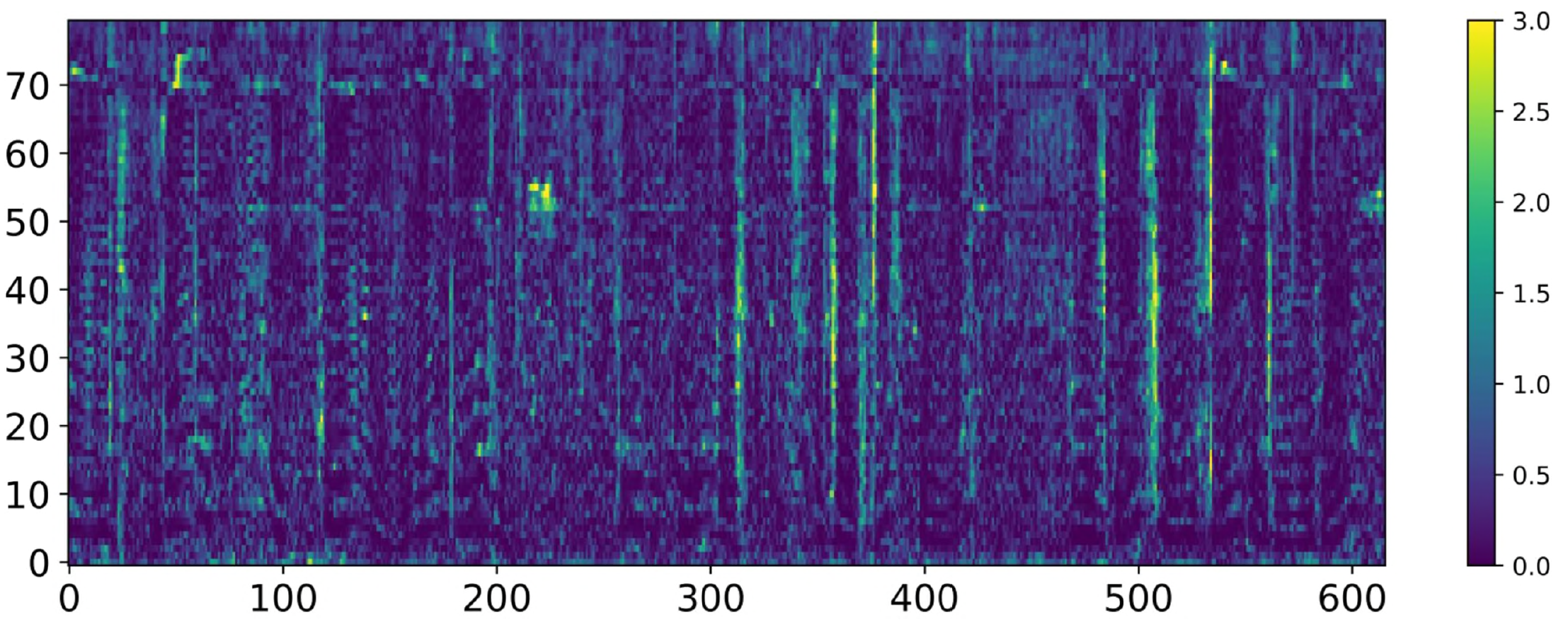}
    \label{fig:dif-waveglow}}
    \caption{Pixel-wise difference in the mel-spectrogram space between generated and ground-truth audio. Fre-GAN reproduces the desired spectral distributions.}
\label{fig:mel-dif}
\end{figure*}

\begin{figure}[!t]
\centering
\hspace{-0.05\columnwidth} 
\subfigure[w/o DWT]{
\includegraphics[width=.495\columnwidth]{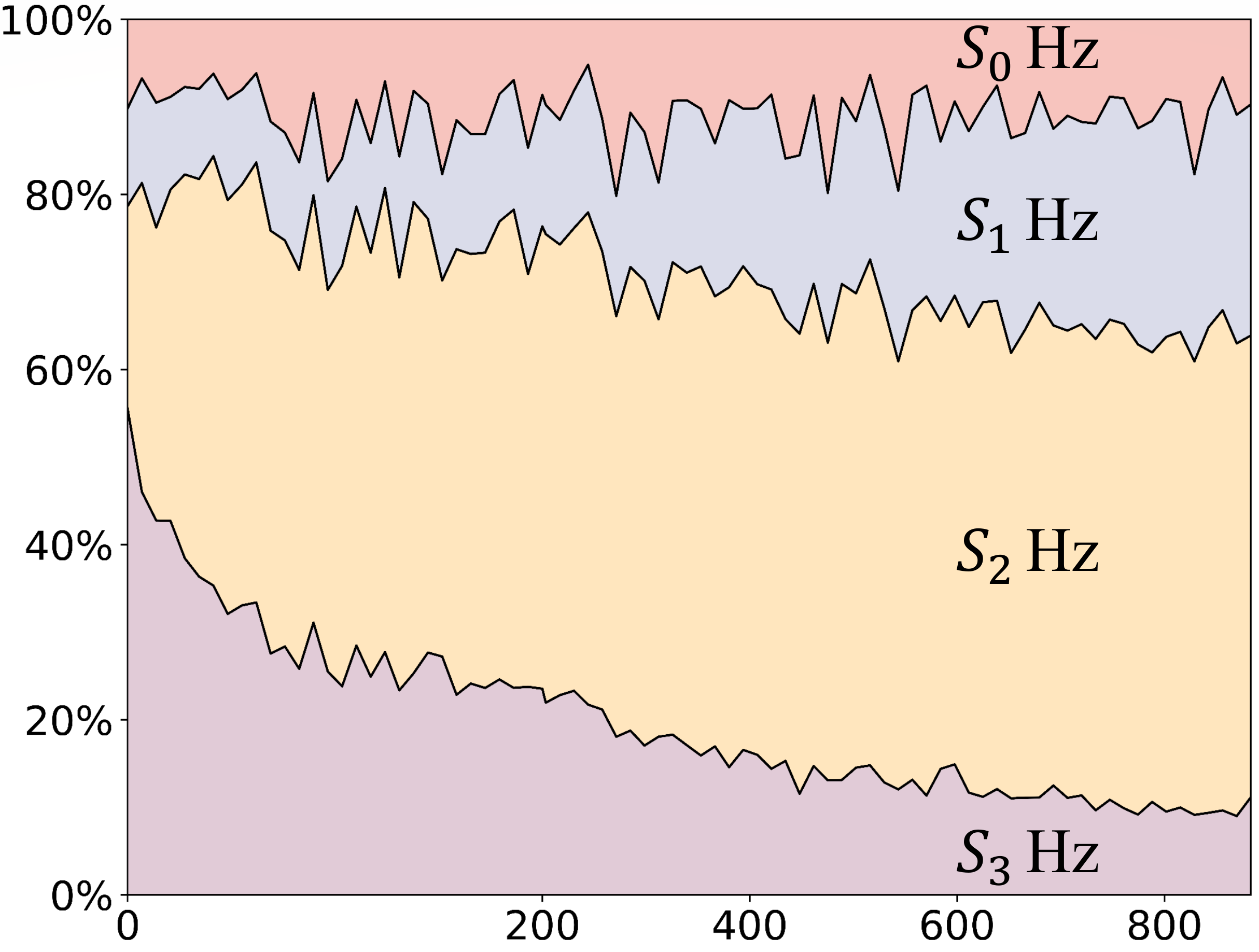}
\label{fig:wo_dwt}}
\hspace{-0.02\columnwidth} 
\subfigure[w/ DWT ]{
\includegraphics[width=.495\columnwidth]{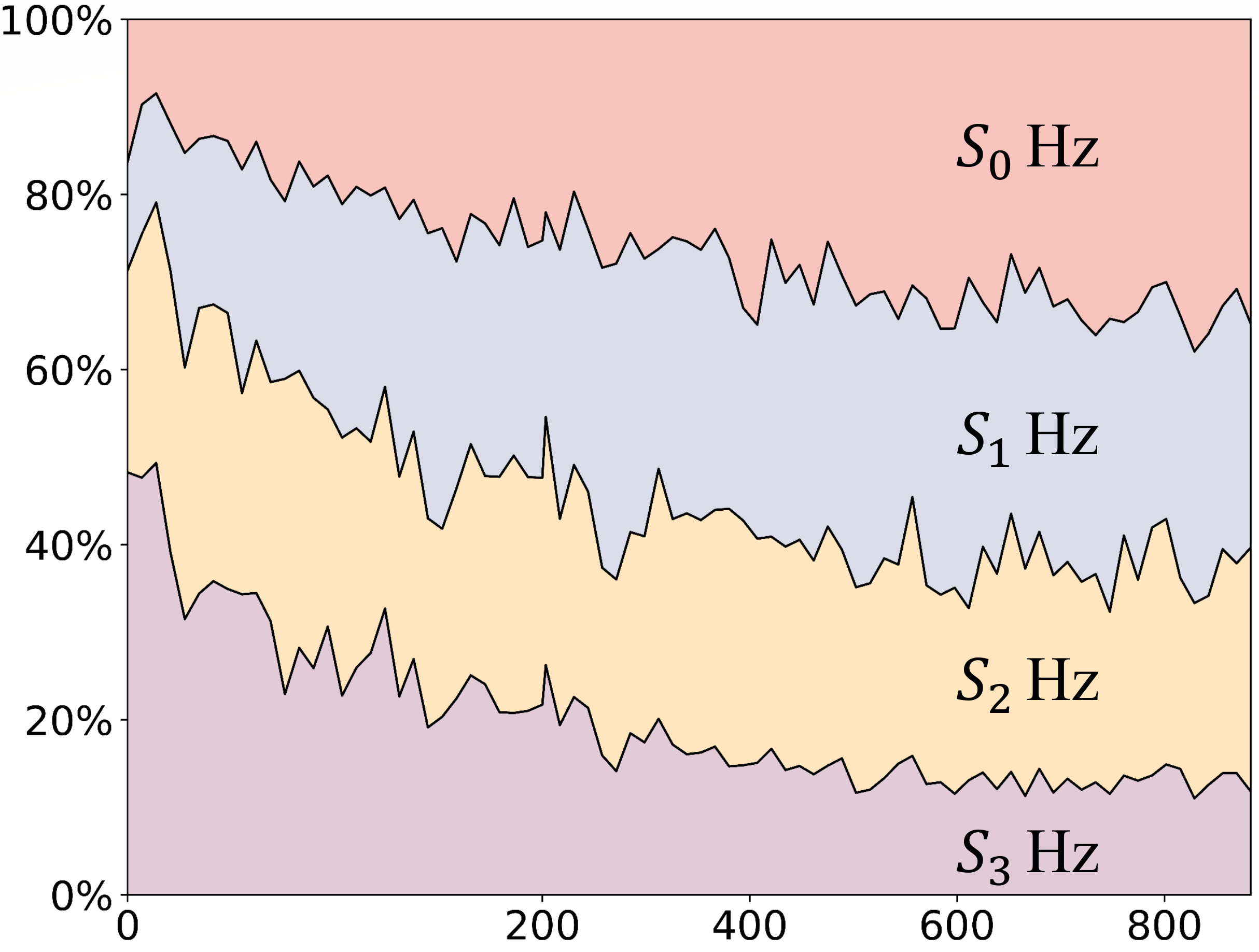}
\label{fig:w_dwt}}
\hspace{-0.05\columnwidth} 
\caption{Contribution of each resolution to the output of the RCG as the training proceeds. The x- and y-axis depicts the training epochs and standard deviations, respectively.}
\label{progressive}
\end{figure}

\subsection{Audio Quality and Inference Speed}
We assessed Fre-GAN on various metrics. As a subjective test, we performed 5-scale MOS tests via Amazon MTurk where at least 20 raters were asked to rate the naturalness of audio. As an objective quality evaluation, we used MCD$_{13}$\cite{kubichek1993mel}, RMSE$_{f0}$\cite{hayashi2017investigation}, and FDSD\cite{binkowski2019high}. 50 and 200 synthesized utterances were used for subjective and objective evaluation, respectively. We also measured the synthesis speed on Intel Xeon Gold 6148 2.40 GHz CPU and a single NVIDIA Titan Xp GPU.

The results are presented in Table~\ref{MOS}. Above all, Fre-GAN outperforms other models in terms of quality. Specifically, Fre-GAN V1 demonstrates similarity to ground-truth audio with a gap of only 0.03 MOS; this implies that the generated audio is highly similar to the real audio. In terms of synthesis speed, Fre-GAN was nearly as fast as HiFi-GAN, and all the variations of Fre-GAN were faster than WaveNet and WaveGlow.

Moreover, we investigated the pixel-wise difference in the mel-spectrogram domain between generated and ground-truth audio, as illustrated in Fig.~\ref{fig:mel-dif}. We observed that the error of the mel-spectrogram was highly reduced in Fre-GAN, which indicates that Fre-GAN generates frequency-consistent audio corresponding to the input mel-spectrogram.

\subsection{Ablation Study}
\label{sec:ablation} 
We performed an ablation study to observe the effect of each Fre-GAN component. Fre-GAN V2 was used as a generator, and each model was trained for $500k$ steps. In Table~\ref{table:ablation}, all the Fre-GAN components contribute to the sample quality. Especially, MOS was largely dropped after replacing DWT with AP, whereas using MPD and MSD (instead of RPD and RSD) shows a relatively small but perceptible degradation. Removing RCG architecture in the generator shows the worst performance in MCD$_{13}$. The absence of mel-spectrogram condition and replacing NN upsampler with transposed convolution also lead to metallic noise and quality degradation.
\begin{table}[!t]
    \caption{MOS, MCD$_{13}$, and FDSD results of ablation study.}
    \centering
    \begin{tabular}{l|c|c|c} \Xhline{3\arrayrulewidth}
        \textbf{Model} &\textbf{MOS} &\textbf{MCD$_{13}$} &\textbf{FDSD}  \\ \hline
            Ground Truth &$4.39\pm{0.03}$ &\textendash  &\textendash \\ \hline 
            Fre-GAN V2 &$\textbf{4.25}\pm\textbf{0.04}$ &$\textbf{1.383}$ &$\textbf{0.209}$ \\ \hline
            \emph{w/o} RCG &$4.14\pm{0.05}$ &$1.602$  &$0.214$ \\
            \emph{w/o} NN upsampler &$4.15\pm{0.05}$ &$1.562$  &$0.238$ \\ 
            \emph{w/o} mel condition &$4.18\pm{0.05}$ &$1.561$ &$0.212$ \\ 
            \emph{w/o} RPD \& RSD &$4.20\pm{0.04}$ &$1.581$  &$0.256$ \\
            \emph{w/o} DWT &$4.12\pm{0.05}$ &$1.592$ &$0.236$ \\ \hline
            HiFi-GAN V2 &$4.08\pm{0.05}$ &$1.793$ &$0.288$ \\
       \Xhline{3\arrayrulewidth}
    \end{tabular}
    \label{table:ablation}
\end{table}

Fig.~\ref{progressive} further verifies the advantages of RCG and DWT. As argued in Sec.~\ref{RCG}, the RCG benefits from progressive learning. To validate this, we quantified the relative importance of multiple waveforms at different resolutions ($S_{0}$, $S_{1}$, $S_{2}$, and $S_{3}$ Hz) by measuring their contribution to the final audio. We calculated the standard deviations of the audio samples as a function of training epochs and normalized the values so that they summed to $100\%$. As shown, the RCG initially focuses on learning low-resolution and then slowly shifts its attention to higher resolutions. One might expect that the highest resolution will be dominant towards the end of training. However, the RCG fails to fully utilize the target resolution when we replace DWT with AP. This implies that replacing DWT with AP washes high-frequency components away, and thus the RCG lacks the incentives to learn high-frequency details.

\section{Conclusions} 
In this paper, we presented Fre-GAN which can synthesize high-fidelity audio with realistic spectra. We observed the inconsistency in frequency space between generated and ground-truth audio, and addressed the problem through the proposed network architecture with a lossless downsampling method. From the experiments, we verified each Fre-GAN component contributes to the sample quality. Additionally, Fre-GAN outperforms standard neural vocoders with only a gap of $0.03$ MOS compared to real audio. For future work, we will apply our proposed method to end-to-end TTS systems to improve the quality of synthesized speech. 

\clearpage
\bibliographystyle{IEEEtran}
\bibliography{mybib}

\end{document}